\documentclass[12pt]{article}
\pdfoutput=1
\usepackage[authoryear]{natbib}
\usepackage{authblk}
\usepackage{array}
\usepackage{multirow}

\usepackage{amsbsy,amsmath,amsthm,latexsym,amssymb}
\usepackage{graphicx,stmaryrd,sectsty}
\usepackage{setspace}
\usepackage[font=small,labelfont=bf]{caption}
\usepackage{verbatim}
\usepackage{graphicx}
\usepackage{caption}
\usepackage{subcaption}
\usepackage{float,color}
\usepackage{url}
\newcommand{\bc}{\begin{center}}
	\newcommand{\ec}{\end{center}}
\newcommand{\bfr}{\begin{flushright}}
	\newcommand{\efr}{\end{flushright}}

\newcommand{\no}{\noindent}
\newcommand{\be}{\begin{enumerate}}
	\newcommand{\ee}{\end{enumerate}}
\newcommand{\bi}{\begin{itemize}}
	\newcommand{\ei}{\end{itemize}}
\newcommand{\bd}{\begin{description}}
	\newcommand{\ed}{\end{description}}
\newcommand{\beq}{\begin{equation}}
	\newcommand{\eeq}{\end{equation}}
\newcommand{\bea}{\begin{eqnarray}}
	\newcommand{\eea}{\end{eqnarray}}

\newcommand{\bfi}{\begin{figure}}
	\newcommand{\efi}{\end{figure}}
\newcommand{\bay}{\begin{array}{l}}
	\newcommand{\eay}{\end{array}}

\newcommand{\cref}[1]{(\ref{#1})}   

\sectionfont{\normalsize}
\subsectionfont{\normalsize}
\subsubsectionfont{\normalsize}
\graphicspath{{./figures/}}
\begin{document}	
	
	
	\begin{titlepage}
		\clearpage\thispagestyle{empty}
		\noindent
		\hrulefill
		\begin{figure}[h!]
			\centering
			\includegraphics[width=2 in]{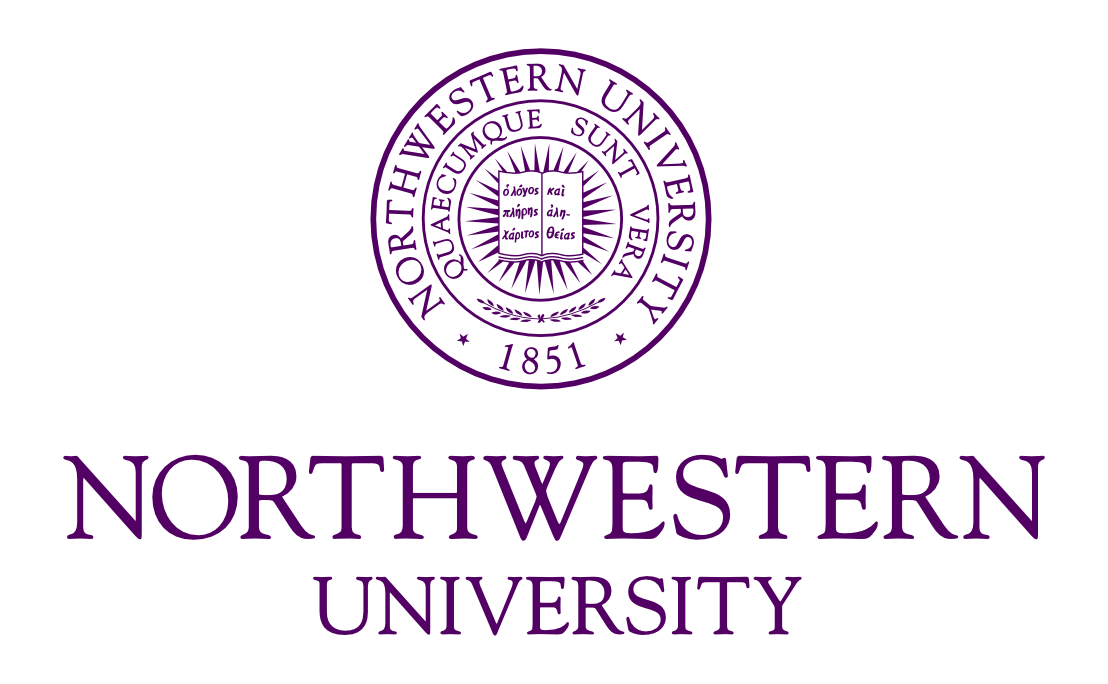}
		\end{figure}
		\begin{center}
			{
				{
					{\bf Center for Sustainable Engineering of Geological and Infrastructure Materials} \\ [0.1in]
					Department of Civil and Environmental Engineering \\ [0.1in]
					McCormick School of Engineering and Applied Science \\ [0.1in]
					Evanston, Illinois 60208, USA
				}
			}
		\end{center} 
		\hrulefill \\ \vskip 2mm
		\vskip 0.5in
		\begin{center}
			{\large {\bf Characterization of Marcellus Shale Fracture Properties through Size Effect Tests and Computations
				}}\\[0.5in]
				{\large {\sc Weixin Li, Zhefei Jin, Gianluca Cusatis}}\\[0.75in]
				{\sf \bf SEGIM INTERNAL REPORT No. 17-08/976E}\\[0.75in]
			\end{center}
			\vskip 5mm
			\noindent {\footnotesize {{\em Submitted to Rock Mechanics and Rock Engineering \hfill August 2017} }}
		\end{titlepage}
		
		\newpage
		\clearpage \pagestyle{plain} \setcounter{page}{1}
		\title{Characterization of Marcellus Shale Fracture Properties through Size Effect Tests and Computations}
		\author[1]{\small Weixin Li\thanks{Email: w.li@u.northwestern.edu}}
		\author[2]{Zhefei Jin \thanks{Email: ZhefeiJin2015@u.northwestern.edu}}
		\author[1,2]{Gianluca Cusatis \thanks{Correspondence to: Gianluca Cusatis. Email:g-cusatis@northwestern.edu}}
		
		\affil[1]{\footnotesize Theoretical and Applied Mechanics, Northwestern University, Evanston, IL 60208, U.S.A.}
		\affil[2]{\footnotesize Department of Civil and Environmental Engineering, Northwestern University, Evanston, IL 60208, U.S.A.}	
		\date{}
		\maketitle
		{\small \no {\bf   Abstract}: 
		Mechanical characterization of shale-like rocks requires understanding the scaling of the measured properties to enable the extrapolation from small scale laboratory tests to field study. In this paper, the size effect of Marcellus shale was analyzed, and the fracture properties were obtained through size effect tests. A number of fracture tests were conducted on Three-Point-Bending (TPB) specimens with increasing size. Test results show that the nominal strength decreases with increasing specimen size, and can be fitted well by Ba\v zant's Size Effect Law (SEL). It is shown that SEL accounts for the effects of both specimen size and geometry, allowing an accurate identification of the initial fracture energy of the material, $G_f$, and the effective Fracture Process Zone (FPZ) length, $c_f$. The obtained fracture properties were verified by the numerical simulations of the investigated specimens using standard Finite Element technique with cohesive model. Significant anisotropy was observed in the fracture properties determined in three principal notch orientations: arrester, divider, and short-transverse. The size effect of the measured structural strength and apparent fracture toughness was discussed. Neither strength-based criterion which neglects size effect, nor classic LEFM which does not account for the finiteness of the FPZ can predict the reported size effect data, and nonlinear fracture mechanics of the quasibrittle type is instead applicable.  
			
		}
		{\small \no {\bf   Keyword}: Marcellus shale; size effect; fracture energy; fracture toughness; fracture process zone}
		

\pagestyle{plain}\thispagestyle{empty}
	

\newpage
\section{Introduction}
In recent years, the study of different aspects of shale-like rocks has surged in popularity as a result of the vital role it plays in various energy-related applications including oil and gas production, subsurface carbon dioxide sequestration, and nuclear waste disposal. Understanding the fundamental mechanical processes in shale formations, as well as their interaction with \emph{in-situ} stress field, pore pressure, and hydraulic loading, is essential to promote industrial innovations such as the development of hydraulic fracturing technique. In particular, the study of crack initiation and propagation in shale-like rocks is of vital importance. However, it is not trivial to characterize the fracture properties of shale to enable the application of fracture mechanics theory in field study, development of numerical tools, and technical design. 

Fracture characterization of shale is usually based on Linear Elastic Fracture Mechanics (LEFM) theory. The mode I fracture toughness, $K_{Ic}$, based on LEFM has been widely used for the characterization of intact rock with respect to its resistance to crack propagation.  \cite{schmidt1977fracture} investigated the fracture toughness of Anvil Point oil shale using three-point-bending specimens with three principal notch orientations being divider, arrester, and short-transverse. The measured $K_{Ic}$ values, varying from 0.3 to 1.1 MPa $\sqrt{\text{m}}$, were found to decrease with an increase in kerogen content and be highest for the divider configuration while lowest for short transverse. Stable crack growth was observed under control of crack opening displacement except for the crack growth perpendicular to the bedding planes (arrester configuration). \cite{chong1987fracture} proposed a Semi-Circular Bend (SCB) specimen subjected to three-point-bending loading for fracture toughness measurement. The $K_{Ic}$ values of Colorado oil shale with divider orientation, determined by using a stress intensity factor method, a compliance method, and a \textit{J}-integral based method, were reported to vary from 0.88 to 1.0 MPa$\sqrt{\text{m}}$ with a change in organic content. In contrast to Schmidt's observation, it was found that the static fracture toughness of organic-rich oil shale is higher than that of lean material. Through Chevron Notched Semicircular Bend (CNSCB) tests, \cite{sierra2010woodford} reported that the fracture toughness of Woodford shale ranging from 0.74 to 1.17 MPa$\sqrt{\text{m}}$ is related to the clay content of the samples. \cite{lee2015interaction} performed SCB tests on Marcellus shale core containing calcite-filled nature fractures (vein). These tests showed that the presence of calcite-filled veins has significant impact on crack propagation paths. For the unfractured samples, $K_{Ic}$ was reported to vary from 0.18 to 0.73 MPa$\sqrt{\text{m}}$ depending on the sample bedding plane orientations. \cite{chandler2016fracture} reported fracture toughness measurements on Mancos shale determined in divider, short-transverse, and arrester configurations, respectively, using a modified short-rod methodology. The highest $K_{Ic}$ value, 0.72 MPa$\sqrt{\text{m}}$ was obtained for crack plane normal to the bedding, and the lowest one, 0.21 MPa$\sqrt{\text{m}}$ for crack plane aligned with the bedding. In addition to conventional fracture tests on notched specimens, some novel testing methodologies, such as scratch tests proposed by Akono \citep{akono2016microscopic,kabirrate}, were also utilized for fracture characterization of various types of shale rocks. Despite the abundance of the fracture experimental data, only limited studies focused on the size and geometry dependence of the shale fracture properties measured from laboratory tests, which, indeed, is non-negligible. For instance, \cite{wang2017experimental} measured the fracture toughness of a shale outcrop in Chongqing, China using SCB and Cracked Chevron-Notched Brazilian Disk (CCNBD) specimens, and noticed that the obtained toughness values from these two methods were different.  

It has been known for some time that fracture toughness based on LEFM applied to laboratory size rock specimens is often underestimated compared to \emph{in-situ} toughness determined from field data \citep{chong1989size,chong1984mechanics}. The measured toughness of various geomaterials was observed to vary with the shape and size of the investigated specimens \citep{ingraffea1984short, kataoka2015size, barpi2012fracture, bocca1989fracture, khan2000effect, wang2017determination,ayatollahi2014size}. For instance, \cite{kataoka2015size} observed that the fracture toughness of Kimachi sandstone increased as the radius of the SCB specimens increases from 12.5 to 150 mm, and converged to a constant value for a radius larger than 70 mm. Indeed, such a size dependency of the measured mechanical properties at laboratory is common in any quasi-brittle material \citep{bavzant1984size}, and is the consequence of material heterogeneity and non-negligible size of the Fracture Process Zone (FPZ). Considering that shale is often regarded as a heterogeneous material, which can be characterized at different length scales \citep{li2016integrated, li2017multiscale}, the size dependence of shale mechanical responses can not be ignored. As a consequence, the fracturing behavior and the energetic size effect associated with the given structural geometry cannot be described by means of classic LEFM. 

Realizing the quasibrittle-type mechanical behavior of shale, one may obtain its fracture properties through size effect testing. The size effect method, originally proposed by Ba\v zant \citep{bavzant1984size,bazant1987determination,bazant1997fracture}, provides an indirect way of measuring the fracture energy, and requires only the knowledge of the peak load, which makes it easier to implement than other methods. In addition, it also provides the material characteristic length of quasibrittle fracture mechanics \citep{cedolin2008identification,cusatis2009cohesive}. The size effect method has been used widely for identification of nonlinear fracture properties of concrete and mortar \citep{bazant1987determination,ba1990determination}, carbon-epoxy composites \citep{bazant1996size,salviato2016experimental}, and rocks such as limestone \citep{bavzant1991identification} and granite \citep{ba1990determination}. In this work, size effect tests were performed on anisotropic Marcellus shale in order to obtain its fracture characteristics. 

\section{Experiments}
\subsection{Material characterization}
The shale material used in the current study was taken from the outcrops of the Marcellus Formation. The blocks are black and compact featured by alternating light and dark layers, as illustrated in Fig. \ref{Fig: specimens}a. Visual inspection shows that the materials are free of surface cracks and voids. The sample can be considered to be dry as the water content by mass measured by following ASTM D2216 is less than 0.2\%. The average mass density is 2558 kg/m$^3$.  

Basic characterization of the sample mechanical properties was conducted, including seismic velocity measurement, direct tension, uniaxial compression, and splitting tests.  Material anisotropy was observed for both seismic velocity, elastic properties, and strengths under tensile and compressive loading conditions. Testing results reveal that the elastic behaviors of the Marcellus shale under study can be described by theory of linear elasticity for transversely isotropic media, with the plane of isotropy coinciding with the plane of sedimentary layering. The five independent elastic constants, $E$, $E'$, $\nu$, $\nu'$, $G$ and $G'$, obtained from uniaxial compression tests are listed in Table \ref{tab:elastic}. $E$,  $\nu$ are the Young's modulus and Poisson's ratio in the plane of isotropy; $E'$ and $\nu'$ are the ones in the plane perpendicular to the isotropy plane; $G'$ is the out-of-plane shear modulus. 

\begin{table}[htbp]
	\centering
	\caption{Elastic properties of Marcellus shale obtained from uniaxial compression tests.}
	\begin{tabular}{lcc}
		\hline
		Description & \multicolumn{1}{l}{Symbol (units)} & \multicolumn{1}{l}{Measured value} \\
		\hline
		In-plane modulus & $E$ (MPa) & 37.7 \\
		In-plane Poisson's ratio & $\nu$ (-) & 0.25 \\
		Out-of-plane modulus & $E'$ (MPa) & 16.1 \\
		Out-of-plane Poisson's ratio & $\nu'$ (-) & 0.35 \\
		Out-of-plane shear modulus & $G'$ (MPa) & 6.9 \\
		\hline
	\end{tabular}%
	\label{tab:elastic}
\end{table}%

\subsection{Specimen preparation}
\begin{figure}
	\begin{center}
		\includegraphics[width=0.8\textwidth]{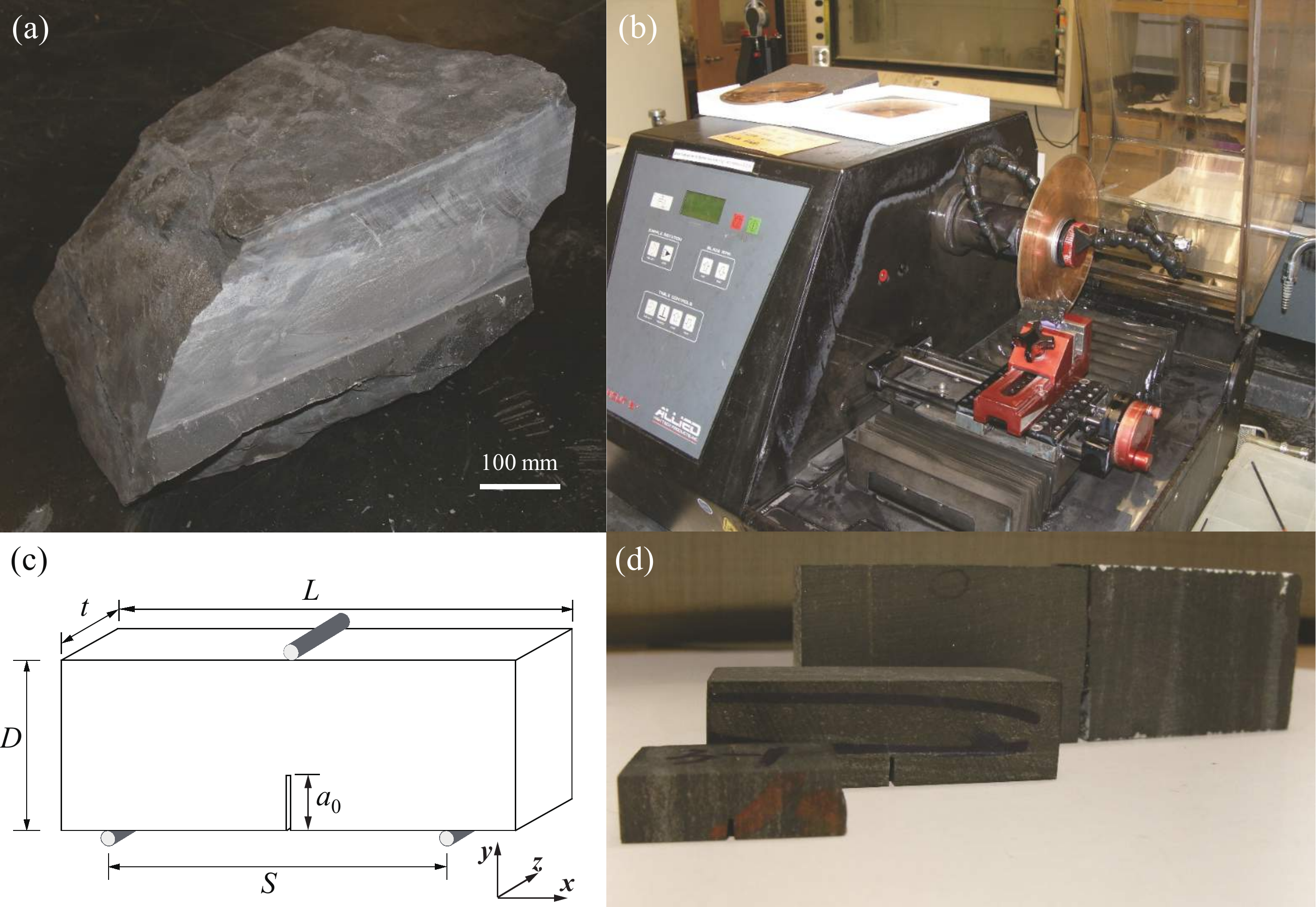}
		\caption{(a) Shale block from Marcellus outcrop. (b) TechCut 5$^{\text{TM}}$ precision sectioning machine. (c) Sketch of three-point-bending (TPB) specimen and loading condition. (d) Typical specimens with increasing size. }
		\label{Fig: specimens}
	\end{center}
\end{figure}

\begin{figure}
	\begin{center}
		\includegraphics[width = 0.8\textwidth]{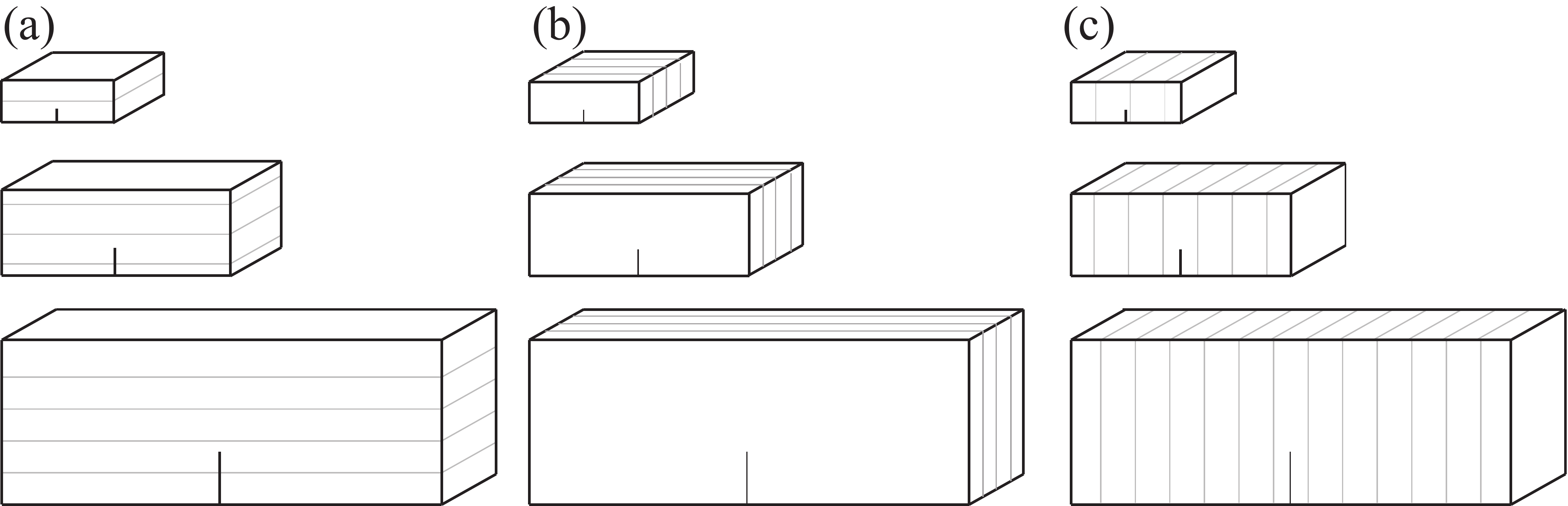}
		\caption{Sketch of the specimens with three principal notch orientations: (a) arrester, (b) divider, (c) short-transverse.}
		\label{Fig: Orien}
	\end{center}
\end{figure}

The large shale block was first cut into small chunks by using a table tile saw with a diamond blade. A TechCut 5$^{\text{TM}}$ precision sectioning machine, as shown in Fig. \ref{Fig: specimens}b, was used to prepare Three-Point-Bending (TPB) specimens with length $L$, depth $D$, thickness $t$, and notches of length $a_0$. A diamond wafering blade with thickness of 0.36 mm was used to machine the notches such that the dimensionless notch length, $\alpha_0 = a_0/D$, was 0.28. Following the pioneering work by \cite{schmidt1977fracture} and \cite{chong1987fracture}, the specimens were made in such a way that the notches were aligned with one of three principal orientations with respect to the isotropy plane, known as arrester, divider, and short-transverse, respectively, as depicted in Fig. \ref{Fig: Orien}. In order to conduct size effect test, specimens with similar geometry and increasing size were prepared for each specimen configuration. Only two dimensional (2D) similarity was treated in this paper, and specimens were designed to be scaled in planar dimensions while kept constant in thickness. Three sizes with ratio of 4:2:1, namely large, medium, and small, were considered. The larger specimens were prepared first. Pieces were collected after the larger ones broke under three-point-bend loading as sketched in Fig. \ref{Fig: specimens}c. In order to reduce machining effort and to minimize the inevitable random scatter of material properties due to shale heterogeneity nature, the medium and small sized specimens were obtained from the collected pieces. The typical TPB specimens with varying sizes are shown in Fig.\ref{Fig: specimens}. The detailed specimen dimensions are listed in Table \ref{tab:geometry}. 

Note that although 2D similarity was prescribed, it is physically difficult to increase or decrease the specimen size while keeping constant geometrical ratios. In particular, the possible largest machining error is in notch length $a_0$, because it is the shortest dimension to machine. In order to check the geometrical similarity condition, the initial notch length was measured after the specimens broke, as listed in the 5th column of Table \ref{tab:geometry}. The table also reported the notch machining error, MAPE$_{a_0}$ which was estimated from the measured and designed values of $\alpha_0$ by means of the formula for Mean Absolute Percentage Error (MAPE) calculation. It turns out that MAPE$_{a_0}$ is not small, resulting that the prepared specimens deviate from the geometric similarity condition. The effect of such geometric imperfection during the machining processes will be discussed later. 

\begin{table}[htbp]
	\centering
	\caption{Geometrical specifications of the TPB specimens under study.}
	\scriptsize
		\begin{tabular}{m{1.2cm}>{\centering}m{1cm}>{\centering}m{1.2cm}>{\centering}m{1cm}>{\centering}m{1.5cm}>{\centering}m{1.5cm}>{\centering}m{1.8cm}>{\centering\arraybackslash}m{1.8cm}}
			\hline
			Type  & Size  & Specimen No. & {Depth, $W$ [mm]} & {Thickness, $t$ [mm]} & {Notch length, $a_0$ [mm]} & Dimensionless notch length, $\alpha_0$ [-] &  \scriptsize{Notch machining error}, MAPE$_{a_0}$ [\%]\\
			\hline
			\multirow{9}[0]{*}{Arrester} & \multirow{3}[0]{*}{Large} & A-L-1 & 25.20 & 14.01 & 7.02  & 0.279 & \multirow{3}[0]{*}{4.82} \\
			&       & A-L-2 & 24.23 & 13.45 & 6.39  & 0.264 &  \\
			&       & A-L-3 & 24.47 & 13.75 & 7.41  & 0.303 &  \\
			& \multirow{3}[0]{*}{Medium} & A-M-1 & 12.52 & 13.81 & 3.47  & 0.277 & \multirow{3}[0]{*}{13.4} \\
			&       & A-M-2 & 12.13 & 14.25 & 4.59  & 0.378 &  \\
			&       & A-M-3 & 12.33 & 13.14 & 3.59  & 0.291 &  \\
			& \multirow{3}[0]{*}{Small} & A-S-1 & 6.13  & 13.98 & 2.16  & 0.352 & \multirow{3}[0]{*}{20.0} \\
			&       & A-S-2 & 5.98  & 14.36 & 2.01  & 0.336 &  \\
			&       & A-S-3 & 6.16  & 12.12 & 1.97  & 0.320 &  \\
			\hline
			\multirow{9}[0]{*}{Divider} & \multirow{3}[0]{*}{Large} & D-L-1 & 25.6  & 12.59 & 7.16  & 0.280 & \multirow{3}[0]{*}{5.28} \\
			&       & D-L-2 & 25.52 & 12.92 & 7.30  & 0.286 &  \\
			&       & D-L-3 & 23.9  & 13.86 & 7.60  & 0.318 &  \\
			& \multirow{3}[0]{*}{Medium} & D-M-1 & 12.7  & 12.86 & 3.55  & 0.280 & \multirow{3}[0]{*}{7.64} \\
			&       & D-M-2 & 12.48 & 13.02 & 3.05  & 0.244 &  \\
			&       & D-M-3 & 11.91 & 14.42 & 3.00  & 0.252 &  \\
			& \multirow{3}[0]{*}{Small} & D-S-1 & 6.41  & 12.70 & 1.65  & 0.257 & \multirow{3}[0]{*}{5.47} \\
			&       & D-S-2 & 6.26  & 13.04 & 1.78  & 0.284 &  \\
			&       & D-S-3 & 6.02  & 14.14 & 1.80  & 0.299 &  \\
			\hline
			\multirow{9}[0]{2cm}{Short-Transverse } & \multirow{3}[0]{*}{Large} & ST-L-1 & 26.12 & 14.10 & 7.04  & 0.270 & \multirow{3}[0]{*}{6.40} \\
			&       & ST-L-2 & 26.2  & 14.00 & 6.40  & 0.244 &  \\
			&       & ST-L-3 & 25.69 & 14.40 & 7.00  & 0.272 &  \\
			& \multirow{3}[0]{*}{Medium} & ST-M-1 & 13.16 & 14.20 & 3.63  & 0.276 & \multirow{3}[0]{*}{2.33} \\
			&       & ST-M-2 & 13.1  & 14.01 & 3.66  & 0.279 &  \\
			&       & ST-M-3 & 12.72 & 14.55 & 3.75  & 0.295 &  \\
			& \multirow{3}[0]{*}{Small} & ST-S-1 & 6.54  & 14.07 & 1.91  & 0.292 & \multirow{3}[0]{*}{4.29} \\
			&       & ST-S-2 & 6.57  & 14.04 & 1.73  & 0.263 &  \\
			&       & ST-S-3 & 6.44  & 14.57 & 1.85  & 0.287 &  \\
			\hline
		\end{tabular}
		\label{tab:geometry}%
	\end{table}%
	
	\subsection{Test description}
\begin{figure}
		\begin{center}
			\includegraphics[width=0.8\textwidth]{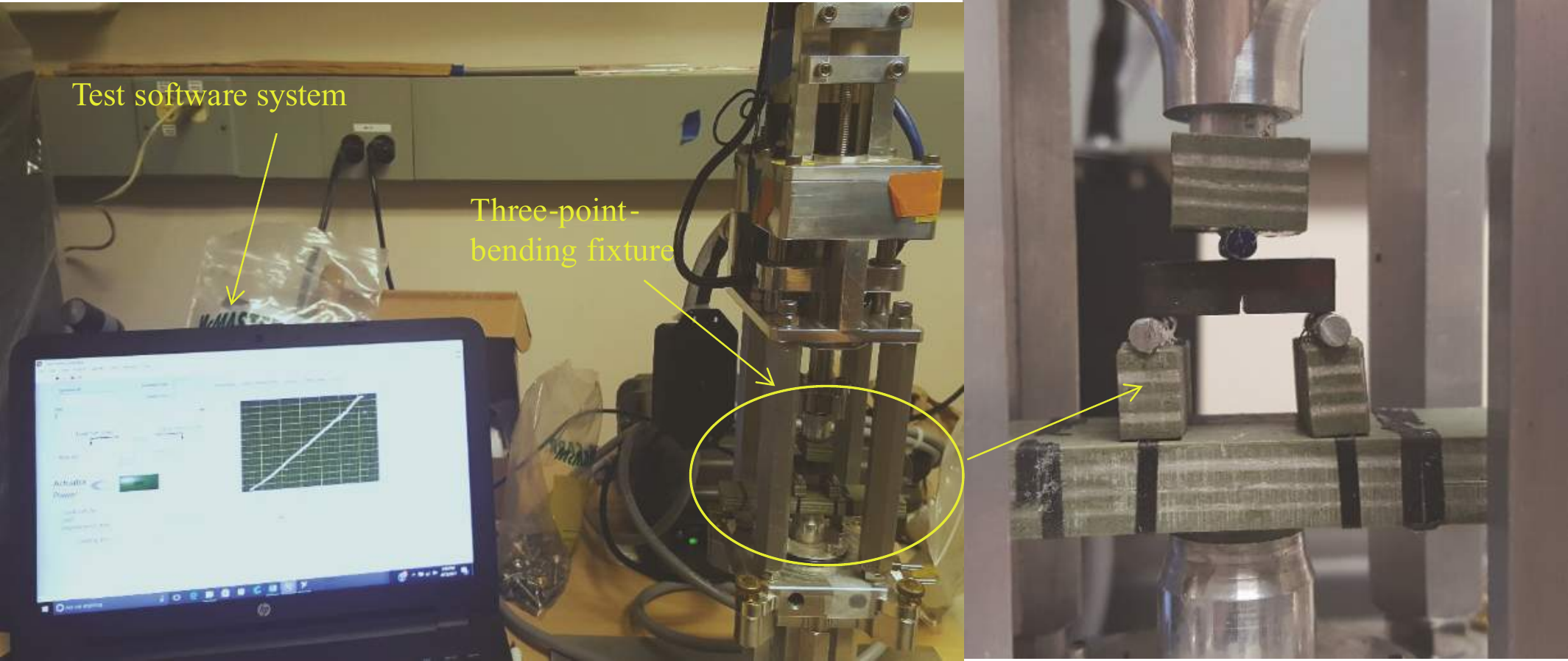}
			\caption{Setup of the loading and data acquisition systems. }
			\label{Fig:setup}
		\end{center}
	\end{figure}

	The prepared TPB specimens were placed on two supporting pins with the support span, $S$, being 74, 37, and 18.5 mm for large, medium, and small size, respectively, and were loaded vertically under symmetrical three-point bending. The tests were conducted under stroke mode on a closed-loop controlled Mini-Tester with a load cell operating in the 200 lb (889.64 N) range. A constant displacement rate of 0.1, 0.05, and 0.025 mm/min was used for large, medium, and small specimens, respectively, to ensure the same strain rate for all investigated specimens. Each test lasted around 5 min to complete. The load-line displacements and loads were recorded during the tests with a system acquisition frequency of 1 Hz. The test configuration is shown in Fig. \ref{Fig:setup}. In total, 27 tests were conducted with three tests for each specimen size and configuration. 
	
	\subsection{Experimental results}
	The recorded load-displacement curves during the experiments can be described by a initial stage with a gentle slope, followed by a segment with linear growth of load, and a sudden drop of load as soon as the peak value was reached. The gentle slope at the initial stage arises from adjusting contact between the specimen and the loading pins. The pre-peak linear segment indicates that no apparent plastic deformation takes place within the tested shale specimens. After reaching the peak load, the load-displacement curves drop instantaneously for all investigated sizes and configurations. As a consequence, the specimens failed and split into two pieces right after the peak load, suggesting a catastrophic (dynamic) failure event. 
	
	Although brittle-like failure behaviors were observed and no measurement of postpeak softening responses was obtained, the brittleness of the material under study needs further investigation. Indeed, stability and controllability of a fracture test depend not only on the material properties, but also on other factors such as machine frame stiffness, control loop feedback mechanism and controller settings, and design of specimen geometry \citep{salviato2016direct}. A possible way to overcome the stability issue of fracture tests is size effect testing which provides a method of calculating the size and geometry independent fracture properties of materials and evaluating specimen brittleness. 
	
	The test results for the notched specimens are summarized in Table \ref{tab:result}. The reported nominal strength, $\sigma_{Nu}$, is defined as the maximum tensile stress at failure based on the unnotched cross section, $\sigma_{Nu} = 1.5(S/D)P_u/Dt$ with $P_u$ being the peak load. The values of the mean and Standard Deviation (SD) for the apparent fracture toughness, $K_{IcA}$, and the apparent fracture energy, $G_{fA}$, calculated from the measured peak load according to LEFM are also reported in the table. It can be seen that not only a variation of the calculated fracture properties with different specimen configurations due to material anisotropy can be observed, but also a variation with specimen size due to significant size effect. One needs to conclude that classic LEFM theory is not sufficient to extract the fracture properties of the material, and strong size and geometry dependency of the test results cannot be ignored.  
	
	\begin{table}[htbp]
		\centering
		\caption{Results of three-point-bending tests on Marcellus shale specimens}
		\scriptsize
			\begin{tabular}{m{1.2cm}>{\centering}m{1cm}>{\centering}m{1.2cm}>{\centering}m{1cm}>{\centering}m{1.5cm}>{\centering}m{2.8cm}>{\centering\arraybackslash}m{2.6cm}}
				\hline
				Type  & Size  & Specimen No. & Peak load $P_u$ [N] & Nominal strength, $\sigma_{Nu}$ [MPa] &  Apparent fracture toughness, $\bar{K}_{IcA} \pm \text{SD}$ [MPa$\sqrt{\text{m}}$] &{Apparent fracture energy, $\quad \bar{G}_{fA} \pm \text{SD}$[N/m]} \\
				\hline
				\multirow{9}[0]{*}{Arrester} & \multirow{3}[0]{*}{Large} & A-L-1 & 503.67 & 6.28  & \multirow{3}[0]{*}{0.851$\pm$0.055} & \multirow{3}[0]{*}{25.344$\pm$3.250} \\
				&       & A-L-2 & 416.60 & 5.86  &       &  \\
				&       & A-L-3 & 429.80 & 5.79  &       &  \\
				& \multirow{3}[0]{*}{Medium} & A-M-1 & 281.00 & 7.20  & \multirow{3}[0]{*}{0.837$\pm$0.143} & \multirow{3}[0]{*}{24.904$\pm$8.671} \\
				&       & A-M-2 & 232.78 & 6.16  &       &  \\
				&       & A-M-3 & 348.17 & 9.67  &       &  \\
				& \multirow{3}[0]{*}{Small} & A-S-1 & 183.02 & 9.67  & \multirow{3}[0]{*}{0.720$\pm$0.093} & \multirow{3}[0]{*}{18.292$\pm$4.793} \\
				&       & A-S-2 & 159.70 & 8.63  &       &  \\
				&       & A-S-3 & 135.98 & 8.20  &       &  \\
				\hline
				\multirow{9}[0]{*}{Divider} & \multirow{3}[0]{*}{Large} & D-L-1 & 498.03 & 6.70  & \multirow{3}[0]{*}{0.967$\pm$0.045} & \multirow{3}[0]{*}{24.815$\pm$2.291} \\
				&       & D-L-2 & 503.39 & 6.64  &       &  \\
				&       & D-L-3 & 413.71 & 5.80  &       &  \\
				& \multirow{3}[0]{*}{Medium} & D-M-1 & 293.00 & 7.84  & \multirow{3}[0]{*}{0.852$\pm$0.033} & \multirow{3}[0]{*}{19.272$\pm$1.482} \\
				&       & D-M-2 & 341.60 & 9.35  &       &  \\
				&       & D-M-3 & 338.65 & 9.19  &       &  \\
				& \multirow{3}[0]{*}{Small} & D-S-1 & 173.31 & 9.22  & \multirow{3}[0]{*}{0.675$\pm$0.050} & \multirow{3}[0]{*}{12.121$\pm$1.810} \\
				&       & D-S-2 & 182.28 & 9.90  &       &  \\
				&       & D-S-3 & 159.32 & 8.63  &       &  \\
				\hline
				\multirow{9}[0]{1.2cm}{Short-Transverse } & \multirow{3}[0]{*}{Large} & ST-L-1 & 488.98 & 5.64  & \multirow{3}[0]{*}{0.820$\pm$0.043} & \multirow{3}[0]{*}{35.913$\pm$3.714} \\
				&       & ST-L-2 & 473.93 & 5.47  &       &  \\
				&       & ST-L-3 & 481.68 & 5.63  &       &  \\
				& \multirow{3}[0]{*}{Medium} & ST-M-1 & 310.19 & 7.00  & \multirow{3}[0]{*}{0.768$\pm$0.010} & \multirow{3}[0]{*}{31.486$\pm$0.819} \\
				&       & ST-M-2 & 309.03 & 7.13  &       &  \\
				&       & ST-M-3 & 292.47 & 6.90  &       &  \\
				& \multirow{3}[0]{*}{Small} & ST-S-1 & 171.78 & 7.92  & \multirow{3}[0]{*}{0.642$\pm$0.049} & \multirow{3}[0]{*}{22.084$\pm$3.437} \\
				&       & ST-S-2 & 177.02 & 8.11  &       &  \\
				&       & ST-S-3 & 194.61 & 8.94  &       &  \\
				\hline
			\end{tabular}
			\label{tab:result}%
		\end{table}%

		\section{Analysis of Experimental Data}
		The size effect test results can be analyzed by means of type II Size Effect Law (SEL) \citep{bazant1997fracture}, which relates the nominal strength, $\sigma_{Nu}$, to the characteristic size of the structure, hereinafter chosen as the specimen depth, $D$. The type II size effect occurs when a large notch or traction-free crack exists at maximum load. The resulting SEL can be derived from an equivalent linear elastic fracture mechanics approach, and bridges the region between strength-based criteria and classic LEFM theory. 
		\subsection{Size effect law for orthotropic materials}
		According to LEFM, Mode I stress intensity factor, $K_I$ of a structure subjected to a nominal stress, $\sigma_N$ can be written as: 
		\begin{equation}\label{Eq:KI}
		K_I = \sigma_N\sqrt{D}k(\alpha) = \sigma_N \sqrt{\pi D \alpha} \xi 
		\end{equation}
		where $k(\alpha)$ and $\xi$ are dimensionless functions. For an orthotropic material, \cite{bao1992role} express $\xi$ as a function of $\alpha$, $\lambda^{1/4}L/D$, and $\rho$, i.e.  $\xi = \xi(\alpha, \lambda^{1/4}L/D, \rho)$, in which $\lambda^{1/4}L/D$ is modified length scale ratio required by orthotropy rescaling, and $\rho$ and $\lambda$ are dimensionless elastic parameters defined as: 
		\begin{equation}
		\rho = \frac{\sqrt{E_x E_y}}{2G_{xy}} - \sqrt{\nu_{xy} \nu_{yx}}, \lambda = \frac{E_x}{E_y}
		\end{equation}
		The Cartesian coordinate system attached to the investigated specimen, as illustrated in Fig. \ref{Fig: specimens}c, is used to define the elastic properties in the equation above, which can be calculated from the measured elastic constants reported in Table \ref{tab:elastic}. 
		
		By relating the energy release rate for orthotropic materials to the stress intensity factor and recalling Eq. \ref{Eq:KI}, one can write $G(\alpha)$ as
		\begin{equation}\label{Eq:G}
		G(\alpha) = \frac{K_I^2}{E^*} = \frac{\sigma_N^2 D}{E^*}g(\alpha)
		\end{equation}
		where
		\begin{subequations}
			\begin{align}
			E^*& = \sqrt{\frac{2E_x E_y \sqrt{\lambda}}{1+\rho}}\\
			g(\alpha)& = k(\alpha)^2 = \pi \alpha \left[\xi(\alpha, \lambda^{1/4}L/D, \rho) \right]^2 \label{Eq:gk}
			\end{align}
		\end{subequations}
		
		Note that Eq. \ref{Eq:G} is similar to the one for isotropic materials except that the effective elastic modulus, $E^*$, is a function of the orthotropic elastic properties, and the dimensionless energy release rate, $g(\alpha)$, accounts for both geometric and elastic effects. Based on equivalent linear elastic fracture mechanics, the crack initiation condition can be written with reference to an equivalent crack length as \citep{bavzant1984size,bazant1997fracture}:
		\begin{equation}
		G\left(\alpha_0 + c_f/D \right) = \frac{\sigma_{Nu}^2 D}{E^*}g(\alpha_0 + c_f/D) = G_f
		\end{equation}
		where $G_f$ and $c_f$ are initial fracture energy and effective FPZ length, respectively, both assumed to be material properties, and failure occurred when the dimensionless equivalent crack length, $\alpha$ equals $\alpha_0+c_f/D$. 
		
		By approximating $g(\alpha_0 + c_f/D)$ with its Taylor series expansion at $\alpha_0$ and retaining only up to the linear term of the expansion, one obtains:
		\begin{equation}\label{Eq:SEL1}
		\sigma_{Nu} = \sqrt{\frac{E^*G_f}{Dg(\alpha_0) + c_f g'(\alpha_0)}}
		\end{equation} 
		This equation relates the nominal strength of structures, $\sigma_{Nu}$, to a characteristic size, $D$, and has the same form as classical Ba\v zant's SEL for the isotropic case. Eq. \ref{Eq:SEL1} can be also recast in the following form:
		\begin{equation}\label{Eq:SEL2}
		\sigma_N = \frac{\sigma_0}{\sqrt{1 + \beta}}
		\end{equation}
		where 
		\begin{subequations}
			\begin{align}
			\sigma_0& = \sqrt{E^*G_f/(c_fg'(\alpha_0))}\\
			\beta& = D/D_0 \label{Eq:beta} \\
			D_0& = c_fg'(\alpha_0)/g(\alpha_0) \label{Eq:D0}
			\end{align}
		\end{subequations}
		Note that Eq. \ref{Eq:SEL2} is endowed with a characteristic length $D_0$ which is usually called the transitional size and is the key to describe the transition from ductile to brittle behavior with increasing structure size. The ratio, $\beta$ of $D$ to $D_0$ is called the brittleness number of a structure, and is capable of characterizing the type of failure regardless of structure geometry \citep{bavzant1991identification}. The brittleness is understood as the proximity to LEFM scaling.  
		
		\subsection{Fitting of experimental data by SEL}
		Providing that the specimens are strictly geometrically similar resulting in constant $\sigma_0$ and $D_0$ for all investigated specimens, Eq. \ref{Eq:SEL2} is typically used for the fitting of experimental data for concrete, composite, and other quasibrittle materials. The geometric requirement can be released if Eq. \ref{Eq:SEL1} is used because the effect of specimen geometry is fully described by $g(\alpha)$ \citep{bazant1997fracture}. It is worth noting that the range of brittleness numbers should be sufficient to obtain statistically acceptable regression results \citep{bavzant1996zero,tang1996variable}.
		
		The fitting of the experimental data based on Eq. \ref{Eq:SEL1} can be conducted through either linear or nonlinear regression approaches \citep{bavzant1996zero,tang1996variable}. Although statistically these two approaches should yield the same results as the number of the tested specimens tends to infinity, the linear approach is preferred and was adopted in this paper because of its simplicity. For the linear regression approach, it is convenient to define the following quantities: 
		\begin{subequations}
			\begin{align}
			X& = \frac{g_0}{g'_0}D,&  Y& = \frac{1}{g'_0\sigma_{Nu}^2} \label{Eq:XY}\\
			A& = \frac{1}{E^*G_f}, & C& = \frac{c_f}{E^*G_f} \label{Eq:AC}
			\end{align}
		\end{subequations} 
		in which $g_0 = g(\alpha_0)$ and $g'_0 = g'(\alpha_0)$. Eq. \ref{Eq:SEL1} can then be expressed in the following form:
		\begin{equation}\label{Eq:LSEL1}
		Y = AX + C
		\end{equation}
		The initial fracture energy, $G_f$, and the effective FPZ length, $c_f$ are directly related to the parameters of the regression equation, $A$ and $B$, which can be estimated from the results of the regression analysis. The dimensionless functions $g(\alpha)$ and $g'(\alpha) = dg/d\alpha$ are needed to complete the fitting of the experimental data. 
		
		By multiplying both sides of Eq. \ref{Eq:LSEL1} by $g'_0/g_0$, one obtains
		\begin{equation}\label{Eq:LSEL2}
		Y' = A' X + C'
		\end{equation}
		in which 
		\begin{subequations}
			\begin{align}
			X'& = D, &  Y'& = \sigma_N^{-2} \label{Eq:XYP}\\
			A'& = \frac{g_0}{E^*G_f}, & C'& = \frac{c_f g'_0}{E^*G_f}  \label{Eq:ACP}
			\end{align}
		\end{subequations} 
		Similarly, the fracture properties can be determined from the parameters of Eq. \ref{Eq:LSEL2}. The regression analysis based on Eq. \ref{Eq:LSEL2} is easier to implement because it shortens mathematical manipulations in calculating the coordinates of data points, and only calculations of $g$ and $g'$ at a fixed $\alpha_0$ for all investigated specimens are needed. However, more machining efforts are desired to guarantee specimen geometrical similarity. On the contrary, the geometric requirement disappears if Eq. \ref{Eq:LSEL1} is used, but more mathematical manipulations are expected.
		
		For convenience, the method based on Eq. \ref{Eq:LSEL1} and \ref{Eq:LSEL2} are referred to as method 1 and 2, respectively, in this paper. These two methods are mathematically equivalent. It is at researchers' discretion to utilize the appropriate one depending upon the accuracy of the scaling in sample preparation. Both of them were applied to analyze the experimental data reported in this paper. A comparison is performed and shown in Section 4. 
		
		\subsection{Calculation of $g(\alpha)$ and $g'(\alpha)$}
		The dimensionless energy release rate, $g(\alpha)$ can be obtained by means of Eq. \ref{Eq:gk} given that the function $ \xi(\alpha, \lambda^{1/4}L/D, \rho)$ is known. \cite{bao1992role} proposed formulas to estimate $\xi$ for a family of notched bars, which, however, do not include the cases of TPB specimens. For lack of closed-form solutions, the function $g(\alpha)$ was calculated numerically by Finite Element Analysis in Abaqus Implicit \citep{abaqus}. The specimens were modeled with 8-node biquadratic plane stress quadrilateral elements (CPS8) while the singularity field at the crack tip was modeled through the quarter element technique \citep{barsoum1974application}. A linear elastic orthotropic constitutive model was used with the material properties obtained from the uniaxial compression tests. The \textit{J}-integral approach was adopted to estimate the energy release rate in the presence of an concentrated force and two supports. The corresponding dimensionless energy release rate can be calculated according to Eq. \ref{Eq:G}. 
		
		In order to obtain $g$ as a function of $\alpha$, the dimensionless stress intensity factor $k(\alpha)$, which is the square root of $g(\alpha)$, i.e. $k(\alpha) = \sqrt{g(\alpha)}$, is assumed to take the following form
		\begin{equation}
		k(\alpha) = \sqrt{\alpha}\frac{p(\alpha)}{(1+2\alpha)(1-\alpha)^{3/2}}\label{Eq:k}
		\end{equation}
		where $p(\alpha)$ is a fourth degree polynomial in $\alpha$. Eq. \ref{Eq:k} was initially formulated for isotropic materials \citep{bazant1997fracture, guinea1998stress}, and is assumed to be also valid for orthotropic  materials. As a result, the function $g(\alpha)$ can be approximated providing that $p(\alpha)$ is known. Accordingly, $g'(\alpha)$ can be calculated as follows:
		\begin{equation}\label{Eq:gp}
		g'(\alpha) = (8 \alpha ^2+1)\frac{p^2(\alpha)}{(\alpha -1)^4 (2 \alpha +1)^3} + \alpha \frac{2p(\alpha)p'(\alpha) }{(1-\alpha )^3 (2 \alpha +1)^2}
		\end{equation}
		The function $p(\alpha)$ is estimated numerically through a polynomial interpolation. 
		
\begin{figure}
			\begin{center}
				\includegraphics[width = 1.0\textwidth]{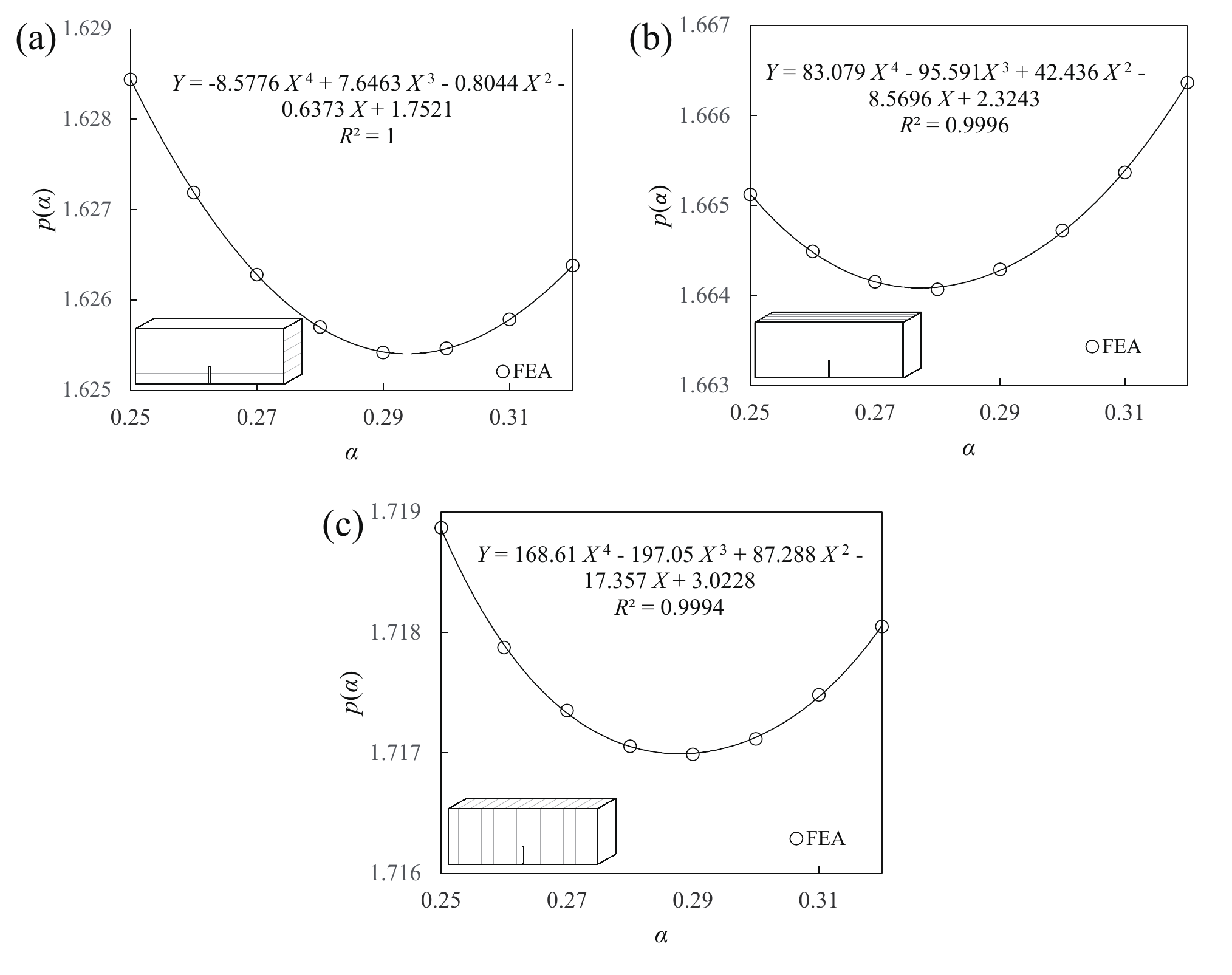}
				\caption{Calculation of function $p(\alpha)$ by fourth degree polynomial interpolation of FEA solutions for the investigated (a) arrester, (b) divider, and (c) short-transverse specimens. }
				\label{Fig: Palpha}
			\end{center}
		\end{figure}

		Various dimensionless crack lengths, from  $\alpha = 0.25$ to $0.32$ in increments of $\Delta \alpha = 0.01$, were considered. The numerically calculated $k$ was used to calculate $p$ at each $\alpha$ by means of Eq. \ref{Eq:k}. As one can note from Fig. \ref{Fig: Palpha} that fourth degree polynomial interpolation provided a very accurate fit of the numerical data for all types of specimens. According to this analysis, one has $p(\alpha) = -8.5776 \alpha^4 + 7.6463 \alpha^3 - 0.8044 \alpha^2 - 0.6373 \alpha + 1.7521$, $\quad83.079 \alpha^4 - 95.591 \alpha^3 + 42.436 \alpha^2 - 8.5696 \alpha + 2.3243$, and $168.61 \alpha^4 - 197.05 \alpha^3 + 87.288 \alpha^2 - 17.357 \alpha + 3.0228$, for arrester, divider, and short-transverse specimens, respectively. The function $g(\alpha)$ can then be calculated according to Eq. \ref{Eq:k} and \ref{Eq:gk}, and $g'(\alpha)$ according to Eq. \ref{Eq:gp}.
		
		\subsection{Identification of fracture properties}
\begin{figure}
			\begin{center}
				\includegraphics[width=1\textwidth]{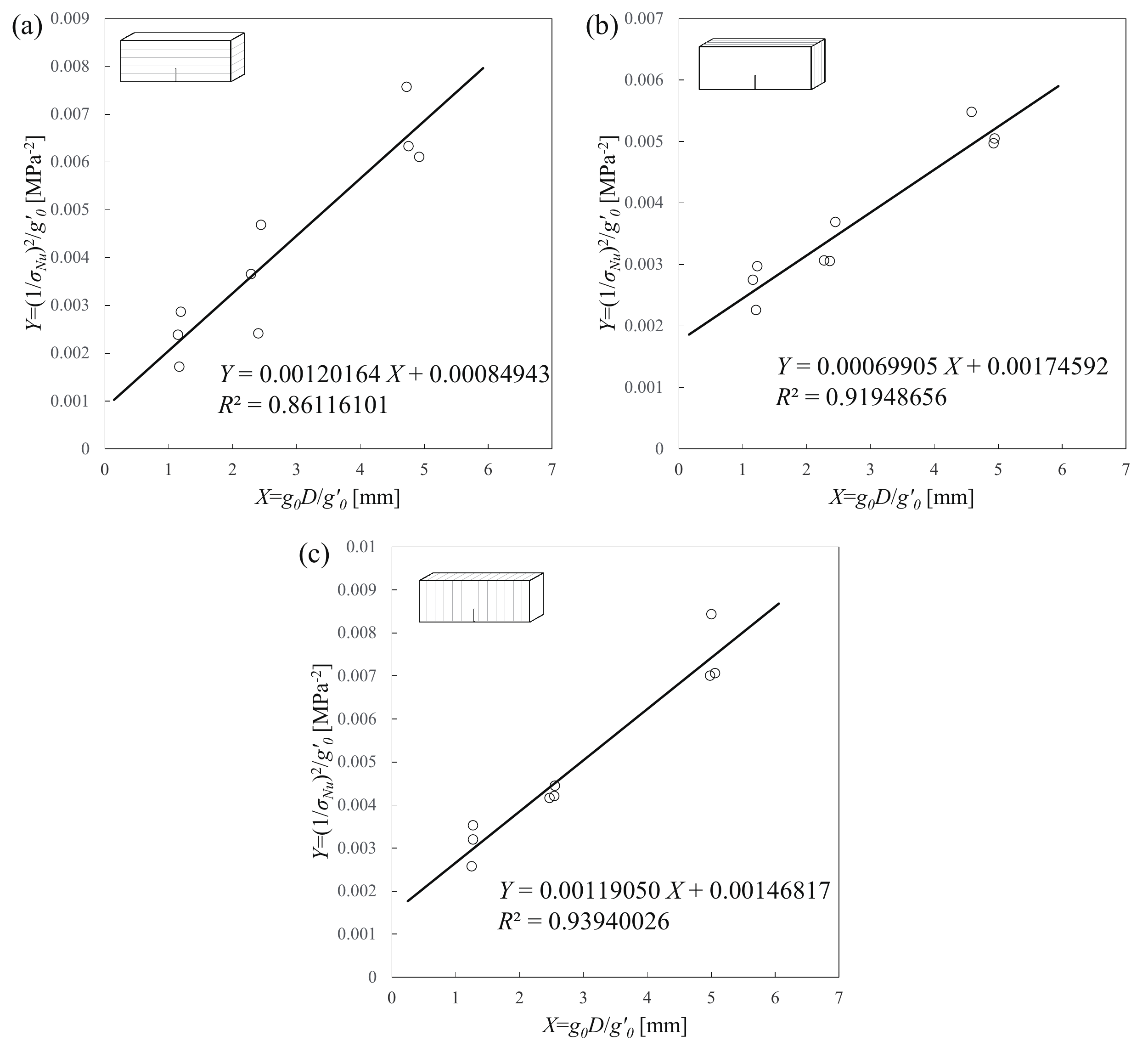}
				\caption{Linear regression analysis based on method 1 considering effect of size and geometry for (a) arrester, (b) divider, and (c) short-transverse specimens}
				\label{Fig:LSEL1}
			\end{center}
		\end{figure}

		Given the calculation of $g$ and $g'$, a linear regression analysis based on Eq. \ref{Eq:LSEL1} (method 1) was conducted by means of ordinary least square method. The results of regression analysis as well as the experimental data are presented in Fig. \ref{Fig:LSEL1}a, b, and c for the arrester, divider, and short-transverse specimens, respectively. The variable $Y$ was plotted against $X$ with $X$ and $Y$ defined in Eq. \ref{Eq:XY}. The regression analysis provided a mean estimate of the parameters $A$ and $C$ with $A$ being the slope and $C$ the intercept of the linear regression equation. In addition, the Standard Error (SE) of the least square estimates can be also provided by the regression formula with the errors assumed to be normally distributed, which were used to quantify the error estimate of the obtained fracture properties. Considering the relation between the fracture properties and the regression parameters (see Eq. \ref{Eq:AC}), one can estimate the means and standard errors of $G_f$ and $c_f$ according to the second-order formulas for the statistics of a function of several random variables \citep{elishakoff1983probabilistic}, which read
		\begin{subequations}
			\begin{align}
			\bar{G}_f& = \frac{1}{E^*\bar{A}}\left( 1+\frac{n \text{SE}^2_A}{\bar{A}^2} \right), \qquad \bar{c}_f = E^* \bar{C} \bar{G}_f \\
			\text{SE}_{G_f}& = \frac{\text{SE}_A}{E^*\bar{A}^2 }, \qquad \text{SE}_{c_f} = \sqrt{\frac{\text{SE}_C^2}{\bar{A}^2} + \frac{\bar{C}^2 \text{SE}^2_A}{\bar{A}^4}}
			\end{align}
		\end{subequations}
		in which $A$ and $C$ are assumed to be statistically independent, and $1+ n \text{SE}^2_A / \bar{A}^2 \approx 1$. The results are reported in Table \ref{tab:fracprop}. The table also provides the values of coefficient of determination (denoted by $R^2$) and Root Mean Squared Error (RMSE) of the estimate based on errors of prediction, both of which quantify the goodness of fit.

\begin{figure}
			\begin{center}
				\includegraphics[width=1\textwidth]{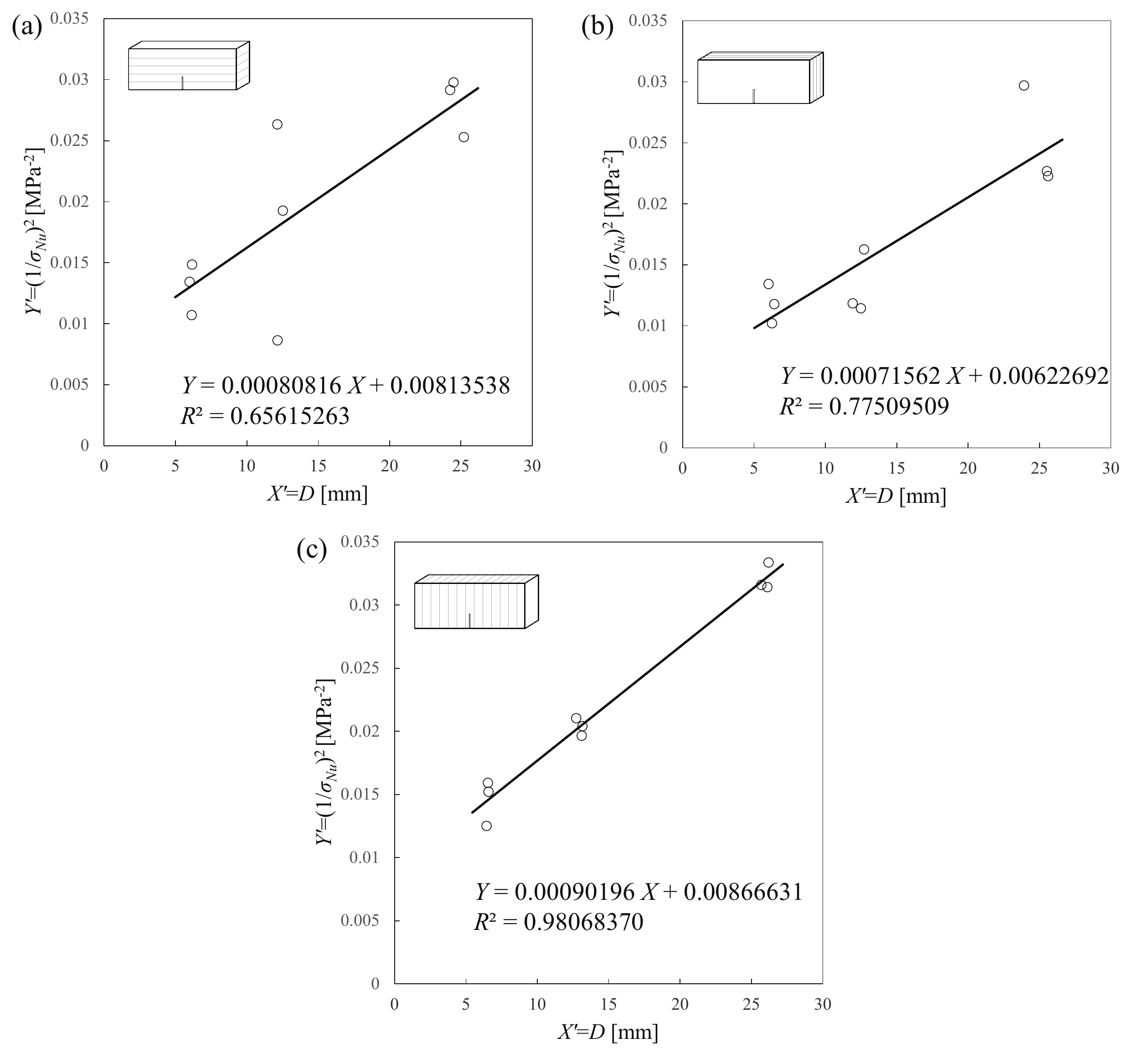}
				\caption{Linear regression analysis based on method 2 considering effect of size for (a) arrester, (b) divider, and (c) short-transverse specimens}
				\label{Fig:LSEL2}
			\end{center}
		\end{figure}

		A regression analysis based on Eq. \ref{Eq:LSEL2} (method 2) was also conducted for comparison. In this case, $g$ and $g'$ were calculated in the same way as described in the above section, but only at a fixed $\alpha_0$ which was calculated by taking the prescribed value of notch length $\alpha_0$ for each sized specimen. The regression results are shown in Fig. \ref{Fig:LSEL2}a, b, and c for the arrester, divider, and short-transverse specimens, in which the variable $Y'$ was plotted against $X'$ with $X'$ and $Y'$ defined in Eq. \ref{Eq:XYP}. Similar to the discussion above, the means and standard errors of $G_f$ and $c_f$ can be obtained from the estimates of the regression parameters according to the least square method and by means of Eq. \ref{Eq:ACP}. The results are reported in Table \ref{tab:fracprop2}.
		
		By comparing the fitting results of method 1 and 2, it is clear to see that better fitting of the experimental data was obtained by means of method 1. Especially, RMSE is found to be almost one order of magnitude smaller for method 1 than the ones for method 2, as listed in Table \ref{tab:fracprop} and \ref{tab:fracprop2}. Given that the random errors in measurement and regression due to material heterogeneity and other random factors are the same in these two cases, the difference of error estimates is mainly due to the machining errors occurred in the notch preparation processes, which, eventually, propagate to calculations of $g$ and $g'$. This can be also proven by comparing the fitting results of the specimens with different configurations. The notch machining error, $\text{MAPE}_{a_0}$ is larger for arrester specimens than the one for short-transverse specimens. Correspondingly, $R^2$ is smaller and RMSE is larger for arrester type specimens. One may also notice that $c_f$ is more susceptible to errors compared to $G_f$ under the normality assumption since SE$_{c_f}$ is around two order of magnitude larger than SE$_{G_f}$ in all cases. Further discussions are based on the fitting results of method 1 since it provided more accurate estimates of the fracture properties of the material. 
		
		\begin{table}[htbp]
			\centering
			\caption{Calculated fracture properties based on size effect method 1 considering effect of size and geometry. }
			\begin{tabular}{lrrrrrr}
				\hline
				Type  & \multicolumn{1}{l}{$R^2$} & \multicolumn{1}{l}{RMSE} & \multicolumn{1}{l}{$G_f$ [N/m]} & \multicolumn{1}{l}{SE$_{Gf}$} & \multicolumn{1}{l}{$c_f$ [mm]} & \multicolumn{1}{l}{SE$_{c_f}$} \\
				\hline
				Arrester & 0.861 & 0.000827 & 29.0  & 0.00440 & 0.731 & 0.492 \\
				Divider & 0.919 & 0.000355 & 37.9  & 0.00424 & 2.99  & 0.452 \\
				Short-Transverse & 0.939 & 0.000534 & 44.8  & 0.00430 & 1.23  & 0.340 \\
				\hline
			\end{tabular}%
			\label{tab:fracprop}%
		\end{table}
		
		\begin{table}[htbp]
			\centering
			\caption{Calculated fracture properties based on size effect method 2 considering effect of size. }
			\begin{tabular}{lrrrrrr}
				\hline
				Type  & \multicolumn{1}{l}{$R^2$} & \multicolumn{1}{l}{RMSE} & \multicolumn{1}{l}{$G_f$ [N/m]} & \multicolumn{1}{l}{SE$_{Gf}$} & \multicolumn{1}{l}{$c_f$ [mm]} & \multicolumn{1}{l}{SE$_{c_f}$} \\
				\hline
				Arrester & 0.656 & 0.00512 & 35.1  & 0.00962 & 0.194 & 1.02 \\
				Divider & 0.775 & 0.00342 & 31.6  & 0.00644 & 1.68  & 0.734 \\
				Short-Transverse & 0.981 & 0.00116 & 53.7  & 0.00285 & 1.86  & 0.203 \\
				\hline
			\end{tabular}%
			\label{tab:fracprop2}%
		\end{table}
		
\begin{figure}
			\begin{center}
				\includegraphics[width = 0.65\textwidth]{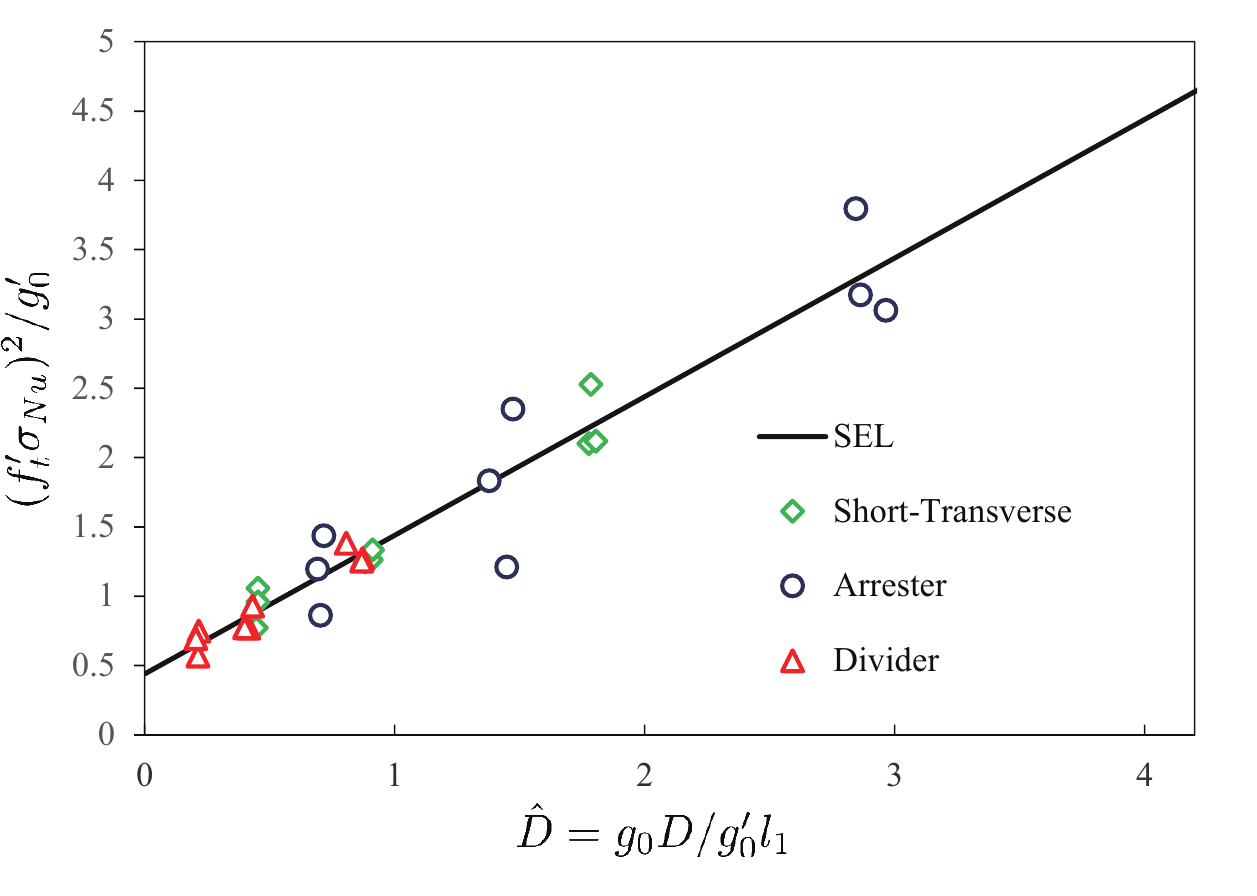}
				\caption{Size effect data in parametric space}
				\label{Fig:Paraplot}
			\end{center}
		\end{figure}

		It is worth noting that there exists a certain optimal size range in which the size effect method gives accurate results. For very small specimens, the two basic assumptions of SEL, vanishing crack tip cohesive stress and constant FPZ length at peak load, may not hold, and the scaling law can be described, instead, by cohesive size effect curves through numerical studies; for very large specimens, the cohesive stress in the FPZ may enter into the tail segment of the cohesive crack law if a bilinear model is used \citep{cusatis2009cohesive}. The optimal size range is often described by a normalized variable $\hat{D} = g_0D/g'_0l_1$ where $l_1 = E^*G_f/f_t'^2$ and $f_t' = \text{tensile strength}$. The lower bound of the optimal $\hat{D}$ was found to be 0.2 \citep{cusatis2009cohesive,yu2009problems}, whereas the upper bound was determined by parameters defining a bilinear cohesive law (total fracture energy $G_F$ etc.). The experimental data presented in this paper is plotted in the parametric space with X-axis represents $\hat{D}$ and Y-axis represents a normalized strength $(f_t' \sigma_{Nu})^2/g_0'$, as shown in Fig. \ref{Fig:Paraplot}. It can be seen that $\hat{D} > 0.2$ for all investigated specimens. As a consequence, the use of SEL to identify the fracture properties is considered to be a valid and simplified alternative to the use of more complicated size effect curves if the cohesive softening law is linear. Although more studies are needed in the case of bilinear or even more complicated softening, it is very likely that only the first linear segment of the cohesive law was approached considering that the range of $\hat{D}$ is less than 1.8 for the divider and short-transverse specimens, and less than 3 for the arrester specimens. Thus, the aforementioned size effect method provided information only on the initial fracture properties in the case of nonlinear cohesive law. 
		
		\subsection{Simulation of size effect tests}
		In order to verify the fracture properties calculated from the size effect tests, numerical analyses were performed on selected specimens of increasing sizes with standard finite element techniques. Finite element models for the selected TPB specimens with the specimen No. listed in Table \ref{tab:simsum} and the dimensions reported in Table \ref{tab:geometry} were built in Abaqus Implicit \citep{abaqus}, and three point bending simulations were performed. The bulk of the discretization was modeled by standard CSP8 elements, and the crack line was modeled by cohesive connections with negligibly small interface thickness of which the behavior is governed by the classic linear traction-separation law. The fracture properties reported in Table \ref{tab:fracprop} and \ref{tab:fracprop2} were used to definite the cohesive surface behavior, and the elastic constants reported in Table \ref{tab:elastic} were used for the elastic orthotropic material model assigned in the bulk region. Note that a fine discretization of the areas adjacent to the crack line is needed in order to capture correctly crack initiation. The element size ahead of the notch tip was kept within the relatively small range of 0.05-0.2 mm, was 1/10 to 1/5 of the FPZ length, and was not scaled upward with specimen size to ensure a similar resolution of the FPZ for all simulations.

		It is also worth noting that the tensile strength, $f_t'$, adopted in the cohesive model was calculated by $f_t' = \sqrt{E^*G_f/l_{1}} $ where $l_{1}$ was related to $c_f$ by $c_f/l_{1} = 0.44$ according to \cite{cusatis2009cohesive}. The tensile strength used in the cohesive model, referred to as fictitious tensile strength, varies from 15 to 20 MPa, which is significantly larger than the tensile strength, ranging from 5 to 10 MPa, measured through the Brazilian split-cylinder tests. The difference is caused by the following two reasons: (1) for typical quasibrittle materials, the tensile strength obtained from the conventional tests, such as Brazilian tests, exhibits a strong size effect \citep{bazant1991size}; (2) the tensile strength identified through the size effect method may be overestimated because the intrinsically nonlinear cohesive crack law is approximated by a linear slope \citep{cusatis2009cohesive}.
		
\begin{figure}
				\begin{center}
					\includegraphics[width=1\textwidth]{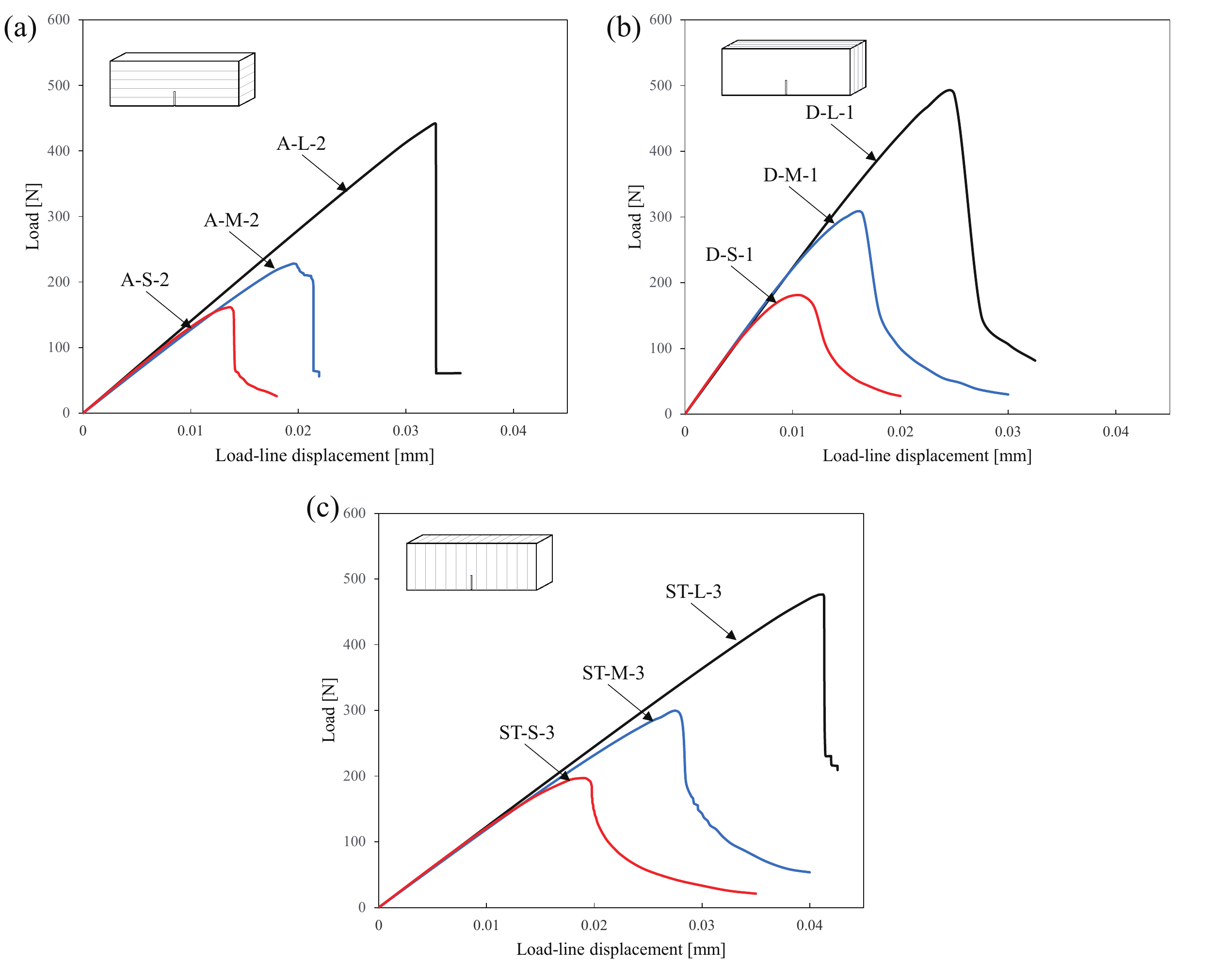}
					\caption{Numerical calculated load-displacement curves of the selected (a) arrester, (b) divider, and (c) short-transverse specimens. }
					\label{Fig:SimLD}
				\end{center}
			\end{figure}

		The numerically calculated peak loads relevant to the investigated specimens were collected and reported in Table \ref{tab:simsum}. The numerical results with the adoption of the fracture properties reported in both Table \ref{tab:fracprop} and \ref{tab:fracprop2},  denoted by Simulation 1 and 2, corresponding to the outcomes of method 1 and 2, respectively, were compared to the experimental data. It can be seen that for all investigated specimens, the predictions with the model calibrated based on the fracture properties obtained via method 1 agree well with experiments, whereas the predicted peak loads were mostly overestimated for the case related to method 2. The simulated load-displacement curves relevant to Simulation 1 in the 5th column of Table \ref{tab:simsum} are also shown in Fig. \ref{Fig:SimLD}.

		\begin{table}[htbp]
			\centering
			\caption{Comparison between experimental and predicted peak load of the selected specimens}
			\scriptsize
				\begin{tabular}{m{1.2cm}>{\centering}m{1.2cm}>{\centering}m{1.8cm}>{\centering}m{1.5cm}>{\centering}m{2cm}>{\centering\arraybackslash}m{2cm}}
					\hline
					\multirow{2}[0]{*}{Type} & \multirow{2}[0]{*}{Size} & \multirow{2}[0]{*}{Specimen No.} & \multicolumn{3}{c}{Peak Load $P_u$ [N]} \\
					\cline{4-6}
					&       &       & {Experiments } & Simulation 1$^a$ (error [\%]) & Simulation 2$^b$ (error [\%]) \\
					\hline
					\multirow{3}[0]{*}{Arrester} & Large & A-L-2 & 416.60 & 442.23 (6.15) & 474.74 (13.96) \\
					& Medium & A-M-2 & 232.78 & 228.18 (1.97) & 303.94 (30.57) \\
					& Small & A-S-2 & 159.70 & 161.47 (1.11) & 184.78 (15.70) \\
					\multirow{3}[0]{*}{Divider} & Large & D-L-1 & 498.03 & 486.32 (2.35) & 514.41 (3.29) \\
					& Medium & D-M-1 & 293.00 & 282.42 (3.61) & 278.17 (5.06) \\
					& Small & D-S-1 & 173.31 & 180.48 (4.13) & 202.27 (16.71) \\
					\multirow{3}[0]{1.2cm}{Short-Transverse} & Large & ST-L-3 & 481.68 & 476.49 (1.08) & 495.29 (2.83) \\
					& Medium & ST-M-3 & 292.47 & 295.49 (1.03) & 306.40 (4.76) \\
					& Small & ST-S-3 & 194.61 & 197.11 (1.29) & 200.68 (3.12) \\
					\hline
					\multicolumn{6}{l}{\scriptsize $^a$ Simulations with the fracture properties reported in Table \ref{tab:fracprop} (method 1)}\\
					\multicolumn{6}{l}{\scriptsize $^b$ Simulations with the fracture properties reported in Table \ref{tab:fracprop2} (method 2)}\\
				\end{tabular}
				\label{tab:simsum}%
			\end{table}%
			
			\section{Discussion}
			\subsection{Size effect on structural strength}
			Size effect on mechanical responses of materials and structures are of concern to many geological engineers as the question is often raised that how the mechanical properties measured in laboratory, which involves millimeter- and centimeter-scale specimens, is applicable to field study of which the typical length of problem ranges from a few meters to a few kilometers. Herein, size effect on two important structural characteristics, strength and fracture toughness, is discussed.

\begin{figure}
				\begin{center}
					\includegraphics[width=0.65\textwidth]{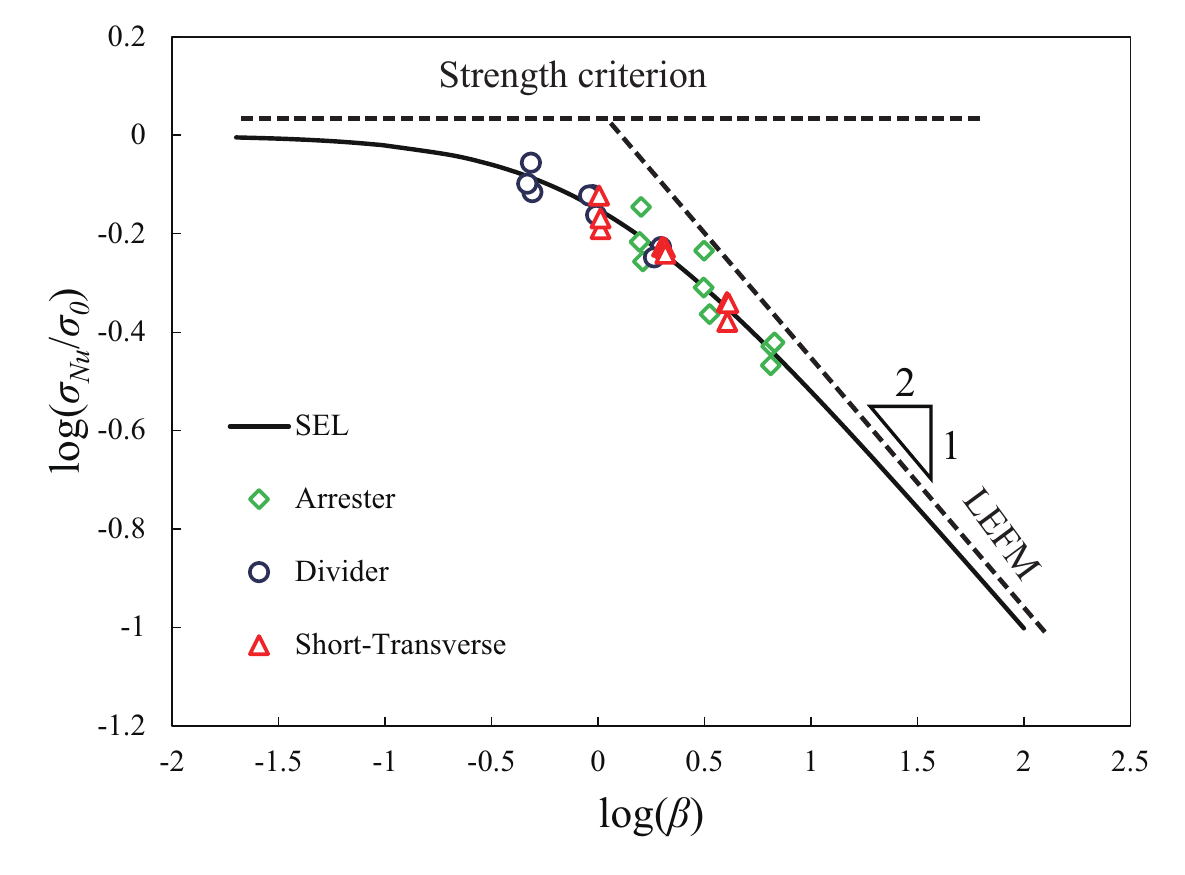}
					\caption{Measured size effect data and plot of size effect curve. }
					\label{Fig:SELsig}
				\end{center}
			\end{figure}

			The measured size effect on structural strength of the investigated specimens are reported in Fig. \ref{Fig:SELsig}  where the normalized strength, $\sigma_{Nu}/\sigma_0$, is plotted as a function of brittleness number, $\beta$, in double logarithmic scale. The brittleness number of each specimen was calculated by $D/D_0$ according to Eq. \ref{Eq:beta}, which, therefore, can be considered as an quantity with regards to the normalized characteristic size. Note that $\beta$ is proportional to the specimen size $D$ and accounts for the effect of the specimen geometry through calculation of $D_0$. Thus, the term \emph{size effect} described by Ba\v zant's SEL hereinafter represents not only the effect of specimen size, but also geometry. The predicted size effect according to SEL (Eq. \ref{Eq:SEL2}) is represented by the solid line in Fig. \ref{Fig:SELsig}. The plot of SEL for structural strength depicts a smooth transition from the strength criterion characterized by a horizontal asymptote when  $\beta \rightarrow 0$, in which no size effect on structural strength is expected, to LEFM by an inclined asymptote of slope -1/2 when  $\beta \rightarrow \infty$, which represents the strongest size effect possible.  
			
			Thanks to the size and shape independent feature, the brittleness number not only relates size effect on structural strength, but also serves as a reliable indicator of the failure type of the tested structure \citep{bazant1997fracture}. For $\beta \rightarrow \infty$, the structure is perfectly brittle and the response follows LEFM, while for $\beta \rightarrow 0$, the structure is perfectly ductile. Quasibrittle structures are those for which $0.1 \leq \beta \leq 10$, in which case the stress and fracture analysis is nonlinear, calling for quasibrittle (cohesive) type models. The $\beta$ values for all investigated specimens in this work fall within the range of 0.4 to 6.8, which are in the transition zone as shown in Fig. \ref{Fig:SELsig}. Therefore, fracture mechanics of the quasibrittle type must be used.  
			
			The relation between specimen size (brittleness number) and failure type can be also observed from the numerically calculated load-displacement curves shown in Fig. \ref{Fig:SimLD}. It can be seen that a vertical drop in load after the peak was observed for all simulated large specimens regardless of notch configuration, suggesting snap-back instability related to brittle failure in the case of greater $\beta$; as the specimen size (also $\beta$) decreases, a trend toward more ductile behavior characterized by gradual post-peak response can be recognized. Note that the term \emph{brittleness} or \emph{quasibrittleness} is relative. If the structure size becomes sufficiently large compared to material inhomogeneities, perfectly brittle behaviors are expected; if the size becomes sufficiently small such that the FPZ extends over the entire structure, the structure becomes perfectly ductile. Furthermore, it is worth pointing out that for all specimens snap-back has to be expected in the experiments due to the deformability of the testing apparatus. This explains why no post-peak was measured in the experiments. The results indicate that a post-peak response is possible to capture in laboratory if specimens are sufficiently small, which, however, would require special miniaturized precision equipments and the tetsting apparatus sufficiently stiff. 
			
			\subsection{Size effect on apparent fracture toughness}
			Apart from the discussion above on structural strength size effect, it is not trivial to also investigate the size dependence of the measured fracture toughness. The term \emph{fracture toughness} is widely used in laboratory and field study, yet there seems to be some confusion between fracture toughness as an unique material characteristic, which does not depend on testing methods, and apparent fracture toughness as a structural property, which, however, is measured at specific specimen size and geometry. The confusion can be clarified through the study of its size and geometry dependency similar to the discussion in \cite{bavzant1991identification}. To avoid confusion, apparent fracture toughness calculated from the measured peak load is denoted by $K_{IcA}$, whereas fracture toughness of material is denoted by $K_{Ic}$.
			
\begin{figure}
				\begin{center}
					\includegraphics[width=0.65\textwidth]{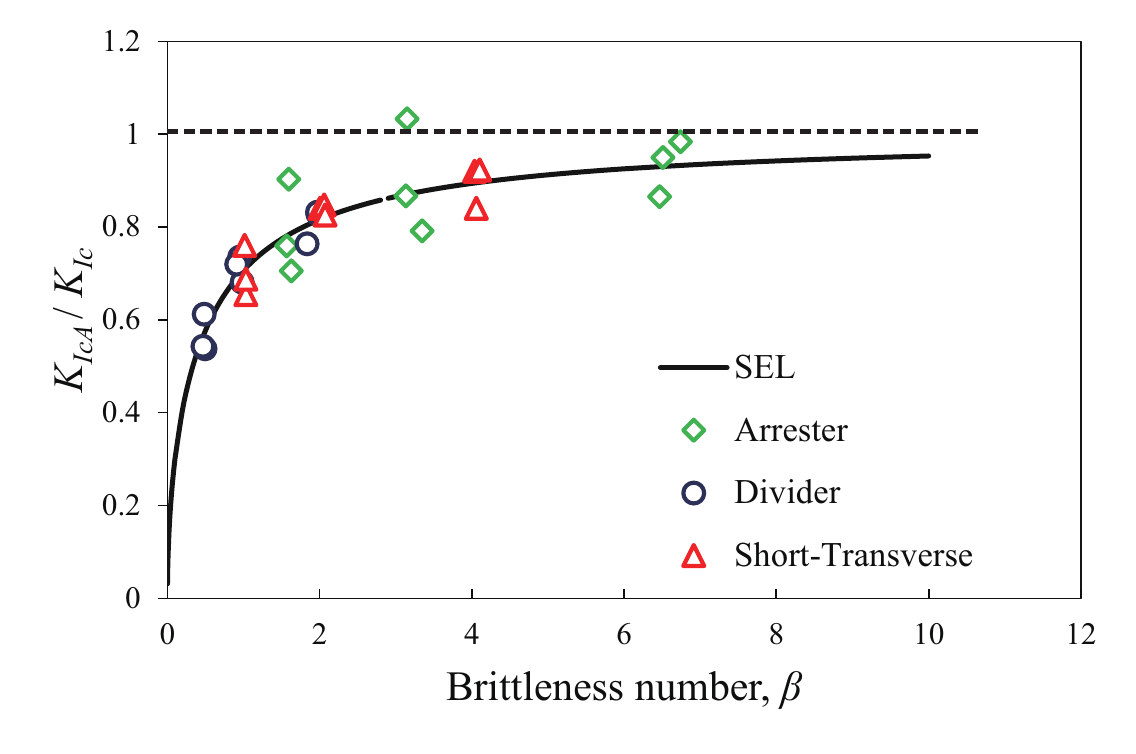}
					\caption{Variation of normalized apparent fracture toughness $K_{IcA}/K_{Ic}$ with brittleness number $\beta$}
					\label{Fig:KICA}
				\end{center}
			\end{figure}

			The normalized apparent fracture toughness of the investigated specimens, $K_{IcA}/K_{Ic}$ is plotted against the corresponding brittleness number, $\beta$, in Fig. \ref{Fig:KICA}. The value of $K_{IcA}$ was calculated according to Eq. \ref{Eq:KI} by letting $\sigma_N = \sigma_{Nu}$, and $K_{Ic}$ was calculated as $K_{Ic} = \sqrt{E^*G_f}$. The properties reported in Table \ref{tab:fracprop} were used to calculate the fracture toughness of each specimen configuration, which yields $K_{Ic} = $ 0.912, 1.20, and 0.917 MPa$\sqrt{\text{m}}$ for arrester, divider, and short-transverse specimens, respectively.  It can be seen from Fig. \ref{Fig:KICA} that for the specimens with a larger brittleness number, a greater apparent fracture toughness was obtained. Specifically, for geometrically scaled specimens of the same type, $K_{IcA}$ increase with specimen size， $D$. This observation is in agreement with the previous fracture tests on different types of rocks. 
			
			The variation of $K_{IcA}$ as a function of $\beta$ can be also predicted by SEL. Substituting Eq. \ref{Eq:SEL1} and \ref{Eq:gk} into Eq. \ref{Eq:KI} and relating $K_{Ic}$ to $G_f$, one can rewrite SEL as
			\begin{equation}
			\frac{K_{IcA}}{K_{Ic}} = \sqrt{\frac{\beta}{1+\beta}}
			\end{equation}
			The equation above is also plotted and represented by the solid line in Fig. \ref{Fig:KICA}. The agreement between the predicted trend and the experimental data is excellent. The ratio of $K_{IcA}$ to $K_{Ic}$ gradually increases as $\beta$ increases and eventually converges to the asymptotic value 1 as $\beta \rightarrow \infty$. In other words, unless the tested specimen is sufficiently large, the fracture toughness of the material cannot be approximated by the apparent one. In practice, $\beta \geq 10$ is required in order to apply classic LEFM and thus to approximate $K_{Ic}$ by $K_{IcA}$. In this case, larger specimens with $D \ge 50$ mm for arrester, $D \ge 125$ mm for divider, and  $D \ge 75$ mm for short-transverse would be needed.
			
			These conclusions ought to be taken into account in various situations relevant to geological engineering design, construction, and operation where a large traction-free crack can grow prior to failure and when extrapolation from small scale laboratory tests to real size structures is needed. In particular, the effect of size becomes extremely important in hydraulic fractures \citep{detournay2016mechanics, chau2016growth, li2017spherocylindrical}. 
			
			\subsection{Anisotropy of fracture properties}
			Significant anisotropy of the measured fracture properties was identified in this work, which may be closely related to crack deflection under complex loading \citep{zeng2017crack,gao2017theoretical}. It was found in previous studies that for the measured fracture toughness of anisotropic shale, the highest value was obtained from divider specimens while the lowest one from short-transverse, i.e. divider $>$ arrester $>$ short-transverse. A similar conclusion can be drawn out from the $K_{IcA}$ measurements reported in this work, as listed in Table \ref{tab:result}, for the large and medium sized specimens. However, this conclusion may be misleading since the comparison was conducted based on the apparent properties measured with specific specimen size and geometry rather that the ``true" properties of materials. In this sense, it is more meaningful to compare the fracture toughness calculated from the size effect tests with the corrections to the size and geometry effects. For the reported $K_{Ic}$ of Marcellus shale, it can be seen that the divider specimens exhibited the highest resistance to Mode I fracture, while the values for the arrester and short-transverse specimens are very close, i.e. divider $>>$ short-transverse $\approx$ arrester. In terms of fracture energy, anisotropy of material deformability also needs to be taken into account considering the relation between $G_f$ and $K_{Ic}$ as shown in Eq. \ref{Eq:G}. From the calculated $G_f$ reported in Table \ref{tab:fracprop}, one can find that short-transverse $>$ divider $>$ arrester.  
			
			Another important characteristic relevant to material fracturability is FPZ length, which, in this work, is quantified by $c_f$, and is strongly related to brittleness of material. As one can conclude from Table \ref{tab:fracprop}, the arrester specimens exhibited the shortest FPZ length while the divider the longest. Considering that $\beta$, which is a measure of structure brittleness, is proportional to $1/c_f$ according to Eq. \ref{Eq:beta} and \ref{Eq:D0}, a material with a smaller $c_f$ tends to be more brittle, and vice versa. As a consequence, relatively more brittle behaviors are expected for the arrester specimens given the same size and geometry. This conclusion agrees with the observation by \cite{chandler2016fracture} that the arrester specimens of Mancos shale exhibited less inelasticity compared to the other ones, and the fracture tests on Anvil Point oil shale conducted by \cite{schmidt1977fracture} which showed that a loss of stability occurred only for the tests on the arrester specimens while not for the other two types of specimens. A comparison of brittleness is also enabled by referring to the simulation results shown in Fig. \ref{Fig:SimLD}. The numerically calculated load-displacement curves for the arrester specimens with medium and small sizes exhibited vertical drop of load after the peak, while the other two types of the specimens showed a certain degree of post-peak responses.
			
			\section{Conclusion}
			Size effect tests were conducted on various TPB specimens of increasing size and different notch configuration to obtain the fracture properties of Marcellus shale in thee principal orientations. The following conclusions can be drawn:
			
			1) Size effect method provides an indirect way of measuring the fracture energy/toughness and effective Fracture Process Zone (FPZ) length, and requires only the knowledge of the peak load. According to this approach, the initial fracture energy, $G_f$, of the investigated material was identified to be from 29.0 to 44.8 N/m depending on notch orientation, the effective FPZ length, $c_f$ from 0.731 to 2.99 mm, and the fracture toughness, $K_{Ic}$ from 0.912 to 1.20 MPa$\sqrt{\text{m}}$. 
			
			2) The Size Effect Law (SEL) proposed by Ba\v zant accounts for the effects of both specimen size and geometry. With the correction to the effect of various notch length, the linear regression results from the fitting of experimental data exhibited less scatters and errors. 
			
			3) The fracture properties calculated via size effect method was verified numerically by means of the standard Finite Element technique with cohesive model. The numerically calculated peak loads using the $G_f$ value estimated by SEL matched the experimental measurements very well. 
			
			4) The experimental investigation shows remarkable size effect on the measured structural strength and apparent fracture toughness, which, however, is often overlooked in the literature on shale fracturing study. Neither strength-based criterion nor classic Linear Elastic Fracture Mechanics (LEFM) theory can predict the size effect data in this paper. On the contrary, the nonlinear fracture mechanics of the quasibrittle type is applicable for fracture characterization of shale in laboratory test. 
			
			5) The brittleness number, $\beta$ introduced by SEL was used to quantify the brittleness of the investigated specimens, which not only dependents on the material characteristics but also the structure size and geometry. 
			
			6) Significant anisotropy in the obtained fracture toughness, $K_{Ic}$, fracture energy, $G_f$, and the effective FPZ length, $c_f$ was observed. 
			\section*{Acknowledgements}
			The authors would like to thank Professor Brad Sageman (Department of Earth and Planetary Sciences, Northwestern University) for providing the Marcellus shale samples used in this study and Professor Giuseppe Buscarnera (Department of Civil and Environmental Engineering, Northwestern University) for his assistance with Mini-Tester. This work also made use of the Materials Characterization and Imaging Facility and the Center for Sustainable Engineering of Geological and Infrastructure Materials (SEGIM) at Northwestern University. 
			
			\bibliography{SEL}
			\bibliographystyle{spbasic}
\end{document}